\documentclass[fleqn,12pt,twoside]{article}
 \usepackage{espcrc1}

\usepackage{graphicx}

\title{Thermal photons and dileptons} 

\author{Charles Gale\address{Department of Physics, McGill University\\
3600 University Street, Montreal, QC, Canada H3A 2T8}}
       
\begin{document}

\maketitle

\begin{abstract}
We discuss the status of a subset of penetrating probes in relativistic nuclear
collisions. Thermal photons and dileptons are considered, as well as the
electromagnetic signature of jets. 
\end{abstract}

\section{Introduction}

The study of penetrating probes constitutes a key aspect of the relativistic
heavy ion program. In the hadronic sector, jet quenching,
for example, 
has been a striking revelation, and has contributed to expose the qualitatively 
different physics that appeared in the transition from the SPS to RHIC. 
In this context, electromagnetic radiation also defines a privileged class of
observables owing mainly to the absence of final state interactions. We
briefly review some recent developments in the measurement of
low- and intermediate-mass lepton pairs, and then address real photon
measurements at RHIC, together with the observed supression in the hard 
parton spectrum.

\section{Lepton pairs}
\subsection{Low invariant masses}
At SPS energies, the measurement of low-mass lepton pairs had previously 
been made by the Helios/3 \cite{helios} and by the CERES \cite{ceres} 
experimental collaborations. As a reminder of the potential of such
measurements for the discovery of new physics, it is useful to write the
emission rate of lepton pairs from a finite-temperature interacting system.
It is \cite{lept}
\begin{eqnarray}
E_+ E_- \frac{d^6 R_{\ell^+ \ell^-}}{d^3 p_+ d^3 p_-} = \frac{2 e^2}{(2
\pi)^6} \frac{n_{\rm B} (E, T)}{M^4} L^{\mu \nu} {\rm Im} \Pi^{\rm R}_{\mu
\nu}
\end{eqnarray}
where $n_{\rm B}$ is a Bose-Einstein distribution function, $L^{\mu \nu}$ is
a lepton tensor, and ${\rm Im} \Pi^{\rm R}_{\mu \nu}$ is the imaginary part
of the in-medium, retarded, finite-temperature self-energy of the photon.
Furthermore, in the nonperturbative sector,  Vector Meson Dominance (VMD)
relates the photon self-energy to the in-medium vector spectral density.  It
is therefore clear that the measurement of the low-mass spectrum of lepton
pairs can reveal pristine features of the interacting many-body system. In
this regard, the situation before this conference was summarized in a
presentation of the CERES collaboration \cite{ceres}, where three theoretical
approaches were shown to be consistent with the data. Those were (i) a many-body
calculation where the in-medium  spectral densities are altered owing to their
coupling with a variety of states accessed through interactions with a hot
and dense hadronic ensemble \cite{RW}; (ii) An effective chiral model, 
where strong precursor effects
already manifest themselves in a shifting of hadronic masses at intermediate
baryonic densities \cite{BR}; (iii)  A simple thermal parametrization of the
quark-antiquark annihilation Born rates, justified by a duality argument
\cite{kam}. Finally and importantly, approaches solely based on vacuum properties have  
difficulty providing an interpretation of the CERES measurements. 

It is probably fair to write that one of the highlights of the Quark Matter
2005 meeting has been the disclosure of the new NA60 data \cite{NA60a} for
In-In collisions at the CERN SPS. The
low-mass dilepton component of this experiment's measurements is of
exceptional quality and statistics \cite{NA60a,NA60b}. The experimental collaboration also has
shown a comparison with their data with the results of approaches (i) and (ii), above. 
On the basis of that comparison, the experimental data strongly suggest that
large mass shifts of the vector mesons are not observed. This same comparison
also shows consistency with the many-body calculations of (i) above. 
An immediate conclusion is the following: NA60 data are now precise enough to
distinguish between different approaches, or at least between some of the 
more extreme scenarios. This is important progress, and
represents a great stride forward. Whether this rules out or not entire
classes of models will remain to be seen, but one point is clear: the
approach shown by the experimental collaboration which is based on in-medium
mass shifts {\it is the same one that was in agreement with CERES results}
\cite{RW}. It
is entirely possible that theoretical refinements are needed in order to
be in line with evolving theoretical paradigms, but the fact remains that any
single theory now has to deal with two separate experimental results. 

The low-mass results do signal an unambiguous many-body effect. To go beyond
this to a state of deeper theoretical understanding is not an easy talk. For
example,  one of the
original aims of this whole experimental program was to isolate a signal 
(precursor or not) of chiral symmetry restoration. This goal has
unfortunately remained elusive, owing largely to the difficulty of directly extracting
an axialvector correlator from relativistic heavy ion data. In the chiral
limit, vector and axialvector correlators are constrained by sum rules of the 
Weinberg-type \cite{wein}, and these may be realized in several different ways
\cite{KS}. The ability to chart a path to correlator degeneracy (in
the Weinberg sense) would definitely represent a significant breakthrough.  
However, the precise data shown at this meeting will surely fuel many
investigations.

\subsection{Intermediate invariant masses}

\begin{figure}[t]
\parbox{8cm}{\includegraphics*[width=8cm]{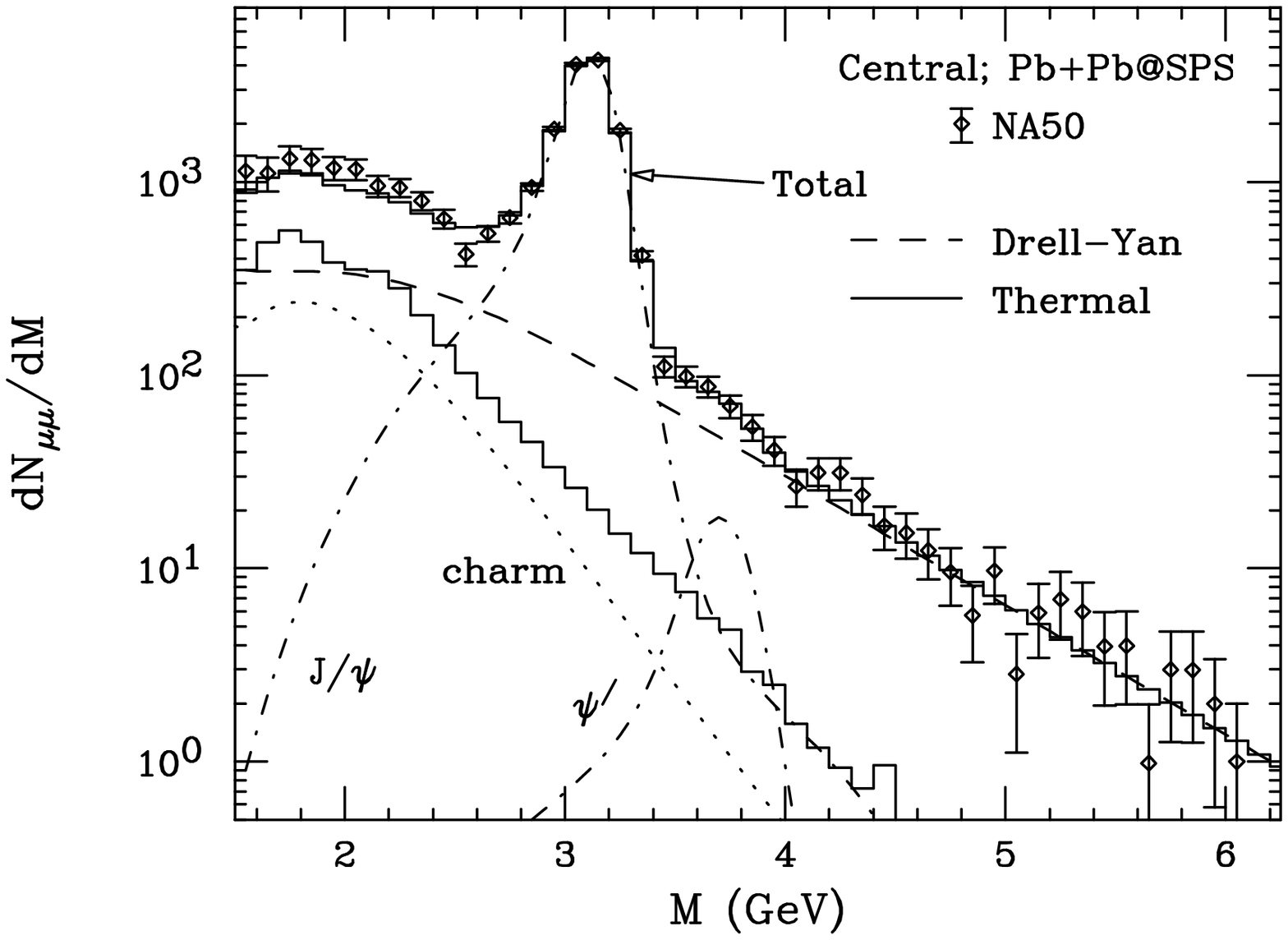}}
\parbox{8cm}{\includegraphics*[width=8cm]{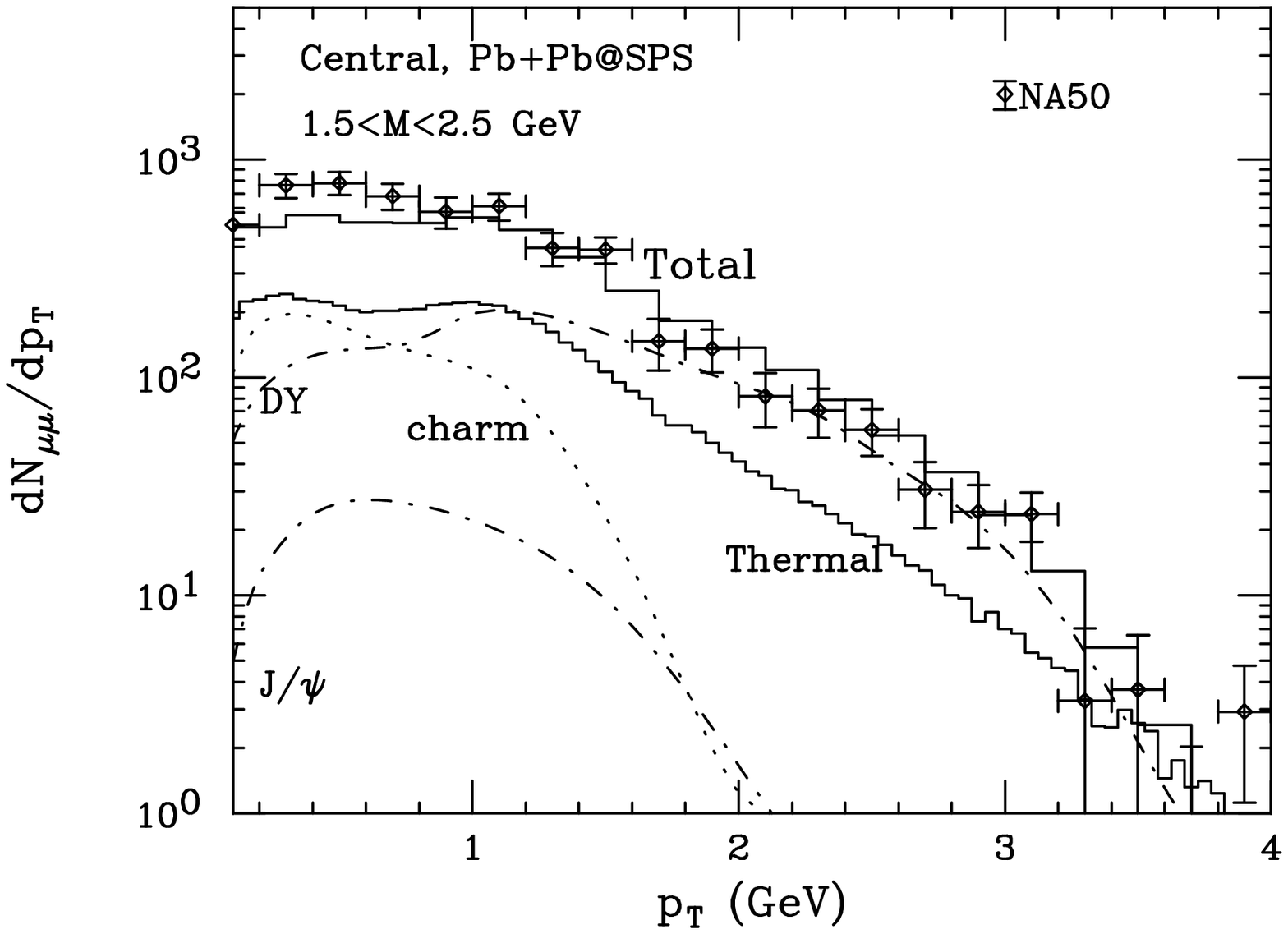}}
\vspace*{-1.0cm}
\caption{The invariant mass and momentum spectra, calculated for dimuon 
pairs. The sources are 
Drell-Yan, correlated charm decay, and thermal (quark-gluon plasma and hadron gas). 
The full curve histogram is the sum of all of those different contributions, 
after correcting for detector acceptance, resolution, and efficiency. 
\label{na50dilep}}
\vspace*{-.5cm}
\end{figure}

The study of lepton pair production at intermediate masses ($m_\phi < M <
m_{J/\psi}$) are especially interesting in the context of searches for the
quark-gluon plasma, as kinematical arguments combined with the original high
temperature of the QCD plasma would designate the intermediate invariant mass
region as a window of opportunity for the observation of plasma radiation
\cite{shu,k2m2}. In this context, considerable interest was generated by
the fact that an excess over sources expected from pA measurements has been
confirmed in the intermediate mass region by the Helios/3 \cite{helios} and 
NA50 \cite{NA50} collaborations. This excess could in turn signal an increase
in $c \bar{c}$ abundances, which would then manifest itself through 
the correlated semileptonic decays of open charm mesons. Alternatively,
thermal lepton pairs need to be ruled out as viable scenario before other explanations
be invoked. Those thermal sources are akin to the ones identified in the low
dilepton mass sector by the many-body calculations. Showing that they also
shine at higher invariant masses would go a long way in providing a 
consistent picture of
electromagnetic radiation in heavy ion collisions. In the theoretical
interpretation of such data, a potential caveat is lurking, and is related  
to the use of effective Lagrangian techniques. These rely on
physical parameters which were essentially all fitted in regions
characterized by soft energy scales, and had mostly to do with strong and
electromagnetic decay widths. When moving over to larger invariant masses,
off-shell effects will set in, making controlled extrapolations a problem
\cite{gaogale}. Fortunately, there exists a wealth of data of the type $e^+
e^- \to$ hadrons, which cover precisely the same range in invariant mass as
the heavy ion data \cite{e+e-}. These data can be analyzed channel-by-channel
and have been used, together with
$\tau$-decay measurements, to construct the vector and axial vector spectral
densities that can be related to the lepton pair spectrum \cite{huang}. 
Summing all kinematically-relevant channels, one arrives at a source which
may be compared to the data via the space-time modeling of the nuclear
collision. The result of one such exercise is shown in Figure
\ref{na50dilep}. 
It is fair to write that the specific values of the temperature evolution, for
example, depend somewhat on the particularities  of the space-time modeling.
However, a fairly robust conclusion still emerges: the intermediate-mass NA50
data does not demand a large contribution from plasma radiation (it is about
20\% here), nor does it need a large enhancement of the initial charm
content. Even though specific details do differ, this conclusion is shared by
other theoretical studies of a similar nature \cite{lg,rs,gkp}. However, in
order to bring this situation to some degree of closure, a direct measurement
of the strangeness component would go a long way. This has now been done and
has been reported at this conference by the NA60 collaboration 
\cite{NA60a,NA60c}. The
first measures taken by the collaboration was to experimentally confirm the
presence of an enhancement in the intermediate-mass continuum in nuclear
collisions, over what is seen in proton-nucleus events. However, with the new
information provided by the muon offset measurement, NA60 can now assert that
the excess is not linked with open charm  enhancement. This new data is
also compatible with the excess that had previously been observed by NA50. The
analysis will proceed further, but it already reveals that the
signal that exceeds what is associated with pA sources increases faster than
linearly with the number of participants. The statistics will improve, 
the 2004 proton-nucleus data will be analyzed and such measurements are
important for the determination of QCD effects beyond leading-twist
\cite{qcd_lt}. It is however 
clear that these results do represent a great leap in our quest for
a complete quantitative understanding of the electromagnetic radiation 
being produced in relativistic nuclear collisions. 

\section{Photons from jet-plasma interactions}

\begin{figure}[htb]
\begin{center}
\includegraphics*[width=7cm,angle=-90]{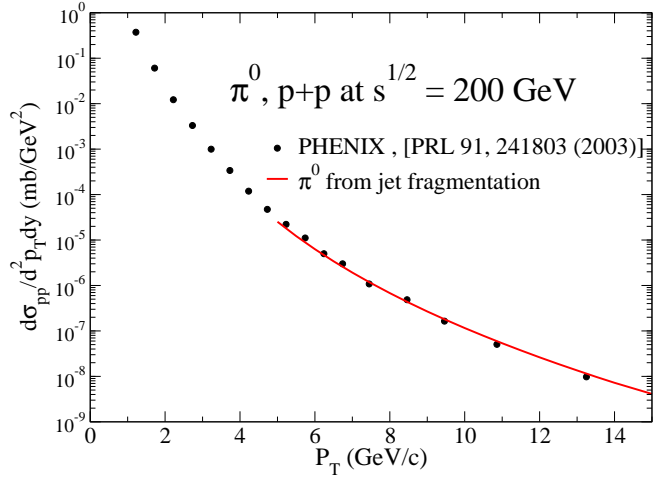}
\vspace*{-1.0cm}
\caption{Neutral pion spectra in pp collisions at RHIC. The data points are
from PHENIX, and the calculated results are from jet fragmentation.} 
\label{pi0}
\end{center}
\vspace*{-1.1cm}
\end{figure}

One of the most striking findings of the RHIC program is the strong
apparent modification of jet characteristics, following a passage through
an interacting, dense medium \cite{xnw}. A compelling theoretical
interpretation of these results is that of jet absorption, 
signaling in effect the existence of a hot and dense partonic phase. Several
models of jet-quenching through gluon bremsstrahlung have been elaborated
\cite{bdmps,ww,vg,kw,z}. Here, we use that of AMY \cite{amy}, because
of its potential to handle consistently jet energy loss and photon emission. 
In this approach, Fokker-Planck equations are solved to obtain the
time-evolution of the initial hard gluon, $P_g (p, t = 0)$ and hard quark plus antiquark
distributions, $P_{q \bar{q}} (p, t = 0)$. The coupled equations are
\begin{eqnarray}
\frac{d P_{q \bar{q}} (p)}{d t} =\int_k P_{q \bar{q}} (p + k) \frac{d
\Gamma_{q g}^q (p+k, k)}{dk dt} - P_{q \bar{q}} (p) \frac{d \Gamma_{q g}^q
(p, k)}{dk dt} + 2 P_g (p +k) \frac{d \Gamma_{q \bar{q}}^g (p+k, k)}{d k dt}
\nonumber \\
\frac{d P_g (p )}{d t} = \int_k P_{q \bar{q}} (p+k) \frac{d \Gamma_{q g}^q
(p+k, p)}{d k dt} 
+ P_g (p+k) \frac{d \Gamma_{g g}^g (p+k, k)}{d k d t}\mbox{\hspace*{3.5cm}} \nonumber \\
\mbox{} - P_g (p) \left( \frac{d \Gamma_{q \bar{q}}^g (p, k)}{dk dt} +
\frac{d \Gamma_{ g g }^g (p, k)}{dk dt} \Theta(2 k - p)\right)
\end{eqnarray}
The kernels $d \Gamma (p, k) / dk dt$ are the transition rates, and they
contain the resummation effects typical of interactions with a thermal medium
\cite{amy}. The solution of the joint equations for the time-evolution of the
parton distribution functions permits the modeling in real time of the
partonic spectra.  The hard parton can then fragment into the different varieties
of observed particles. Up to suppressed corrections, the cross
section for produced pions in nucleon-nucleon collisions can be written 
in a factorized form as
\begin{eqnarray}
 \frac{d^3 \sigma_{\rm pp}}{d^2 p_\bot d y} = \sum_{a, b, c, d} \int d x_a d x_b\, g(
x_a, Q) g(x_b, Q) K_{\rm jet} \frac{d \sigma^{a + b \to c +d}}{d t}
\frac{1}{\pi z} D_{\pi^0 / c} (z, Q')\ ,
\end{eqnarray}
where $g(x, Q)$ is the parton distribution function in a nucleon, $D_{\pi^0 /
c}$ is the pion fragmentation function, $d \sigma^{a b \to c d} /d t$ is the
parton-parton cross section at leading order, and $K_{\rm jet}$ accounts for
higher order effects. Here, ``jet'' essentially means a fast parton with $p_T
\gg$ 1 GeV. This procedure, with the factorization scale $(Q)$ and the
fragmentation scale $(Q')$ set equal to $p_T$, the CTEQ5 parton
distribution functions, and $K_{\rm jet} \sim 1.7$, does a very good job  of
reproducing the measured $\pi^0$ above $p_T \approx $ 5 GeV \cite{tgjm}, in
nucleon-nucleon collisions at RHIC. This
is shown in Figure \ref{pi0}. 

To obtain the high $p_T$ $\pi^0$ cross section in AA collisions, the pp
calculation must be modified in two important ways. First, the parton
distribution function of a nucleus differs from that of a proton:
\begin{eqnarray}
g_{\rm A} (x_a, Q) = g(x_a, Q) R_{\rm A} (x_a, Q)
\end{eqnarray}
The nuclear modification factor of the structure function $R_{\rm A}$ takes
into account shadowing and anti-shadowing \cite{Esk}. Also, the hard parton
looses energy between the initial hard scattering and its hadronization, and
this information in contained in the time-evolution of the entire partonic
profile we have described earlier. The $\pi^0$ spectrum obtained in AA
collisions is then
\begin{eqnarray}
\frac{d^3 N_{\rm AA}}{d y d^2 p_T} = \frac{\langle N_{\rm coll}\rangle}{\sigma_{\rm
in}} \sum_{a,b,c,d} \int dx_a dx_b g_{\rm A} (x_a, Q) g_{\rm A} (x_b, Q)
K_{\rm jet} \frac{d \sigma^{a + b \to c + d}}{d t} \frac{\tilde{D}_{\pi^0 /a}
(z, Q)}{\pi z}
\end{eqnarray}
where $\langle N_{\rm coll} \rangle$ is the average number of binary collisions, 
$\sigma_{\rm in}$ is the inelastic nucleon-nucleon cross section, and the
medium-modified fragmentation function is 
\begin{eqnarray}
\tilde{D}_{\pi^0 /c} (z, Q) = \int d^2 r_\bot {\cal P} ({\bf r_\bot})
\tilde{D}_{\pi^0/c} (z, Q, {\bf r_\bot}, {\bf n} )
\end{eqnarray}
where ${\cal P}$ takes into account the geometry of the emitting source, and 
\begin{eqnarray}
\tilde{D}_{\pi^0 / c} (z, Q, {\bf r}, {\bf n}) = \int dp_f\, \frac{z'}{z}
\left(P_{q \bar{q}/c} (p_f; p_i) D_{\pi^0/q} (z', Q) + P_{g/c}(p_f; p_i) D_{\pi^0 / g}
(z', Q) \right)
\end{eqnarray}
where $z = p_T/p_i$, and $z' = p_T/p_f$. Note also that $P_{q \bar{q} /c}
(p_f ; p_i)$ and $P_{g/c} (p_f; p_i)$ are the solutions to the
Fokker-Planck equation and represent the probability to get a given parton
with final momentum $p_f$, given that the initial configuration is a particle
of type $c$ and momentum $p_i$. Information on the initial temperature
sensitivity, on
the geometry and other details is in Ref. \cite{tgjm}. 

A quantitative measure of in-medium modifications is contained in the
so-called $R_{AA}$ profile
\begin{eqnarray}
R_{AA} = \frac{ \sigma_{\rm in} d^3 N_{AA} / d y d^2 p_T}{\langle N_{coll} \rangle 
d^3 \sigma_{pp} / d y d^2 p_T}\ ,
\end{eqnarray}
when plotted as a function of the transverse momentum. Clearly, if a
nucleus-nucleus collision is nothing but a superposition of nucleon-nucleon
collisions, then $R_{AA}$ should be unity.  This variable is now available
for a variety of particles, over a wide range of transverse momenta. 
\begin{figure}[htb]
\begin{center}
\includegraphics*[width=7cm,angle=90]{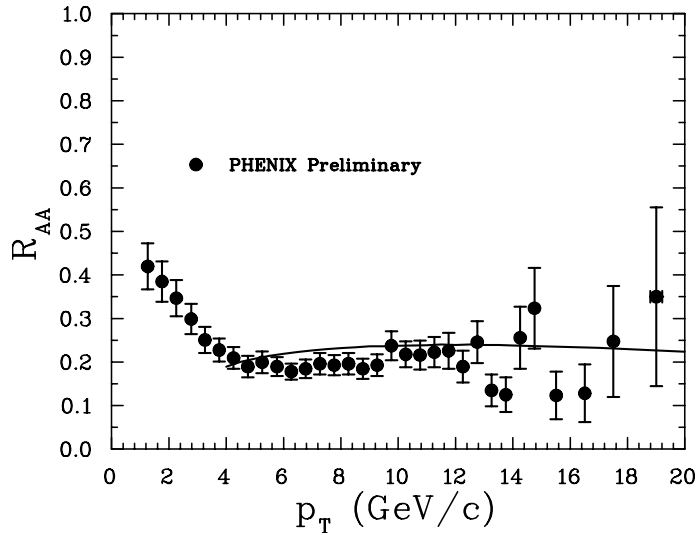}
\vspace*{-1.0cm}
\caption{The ratio $R_{AA}$ for $\pi^0$, as a function of transverse momentum.
The full line is the calculation described in this text, and in \cite{tgjm}, 
with $\alpha_{\rm s}$ = 0.34. 
The initial temperature is 370 MeV, and is consistent with the study in
\protect\cite{trg}.  
Data are from \protect\cite{cole}.  \label{RAA}}
\end{center}
\vspace*{-1.1cm}
\end{figure}
Figure \ref{RAA} reveals new preliminary data, where the behavior of 
$R_{\rm AA}$ over an 
impressive range of almost 20 GeV/c is shown. 
Also there, is the result of 
a calculation with the formalism described
here.  In this approach, the strong coupling constant, $\alpha_{\rm s}$, is a
free parameter. In his plot, a value of $\alpha_{\rm s} = 0.34$ yields a good
fit. This value is kept
fixed for the rest of this work. Further note there are uncertainties
associated with the determination of $<N_{\rm coll}>$, as well as inherent
systematic measurement uncertainties. This combination roughly is of the
order of 10\%. The apparent agreement (for $p_T \geq 4$
GeV)  is at least
superficially satisfying, even if many questions remain on details of the
opacity of the medium, and on the actual sensitivity of this variable on bulk
properties. The answer to those questions is at present scheme-dependent. 
With data of this quality however, the next meeting in this
series is bound to see progress on these issues. Nevertheless, the flatness
of these spectra appears a robust feature. The softer part of the spectrum is
not reproduced by a QCD fragmentation framework and requires 
additional ingredients possibly related to parton recombination \cite{denes}.

\begin{figure}[htb]
\begin{center}
\parbox{6cm}{\hspace*{-1cm}\includegraphics*[width=5.5cm,angle=-90]{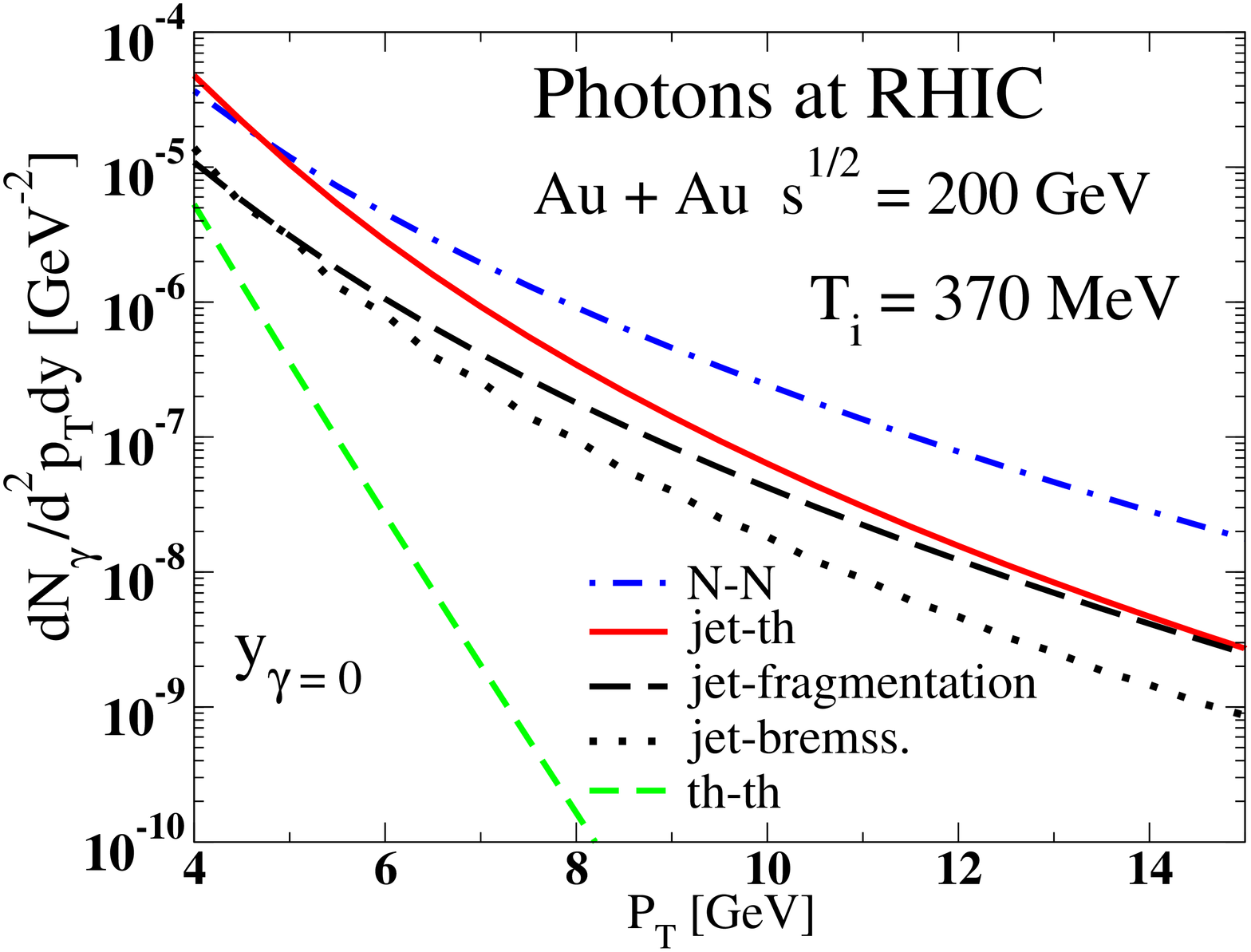}}
\parbox{6cm}{\hspace*{0.3cm}\includegraphics*[width=5.5cm,angle=-90]{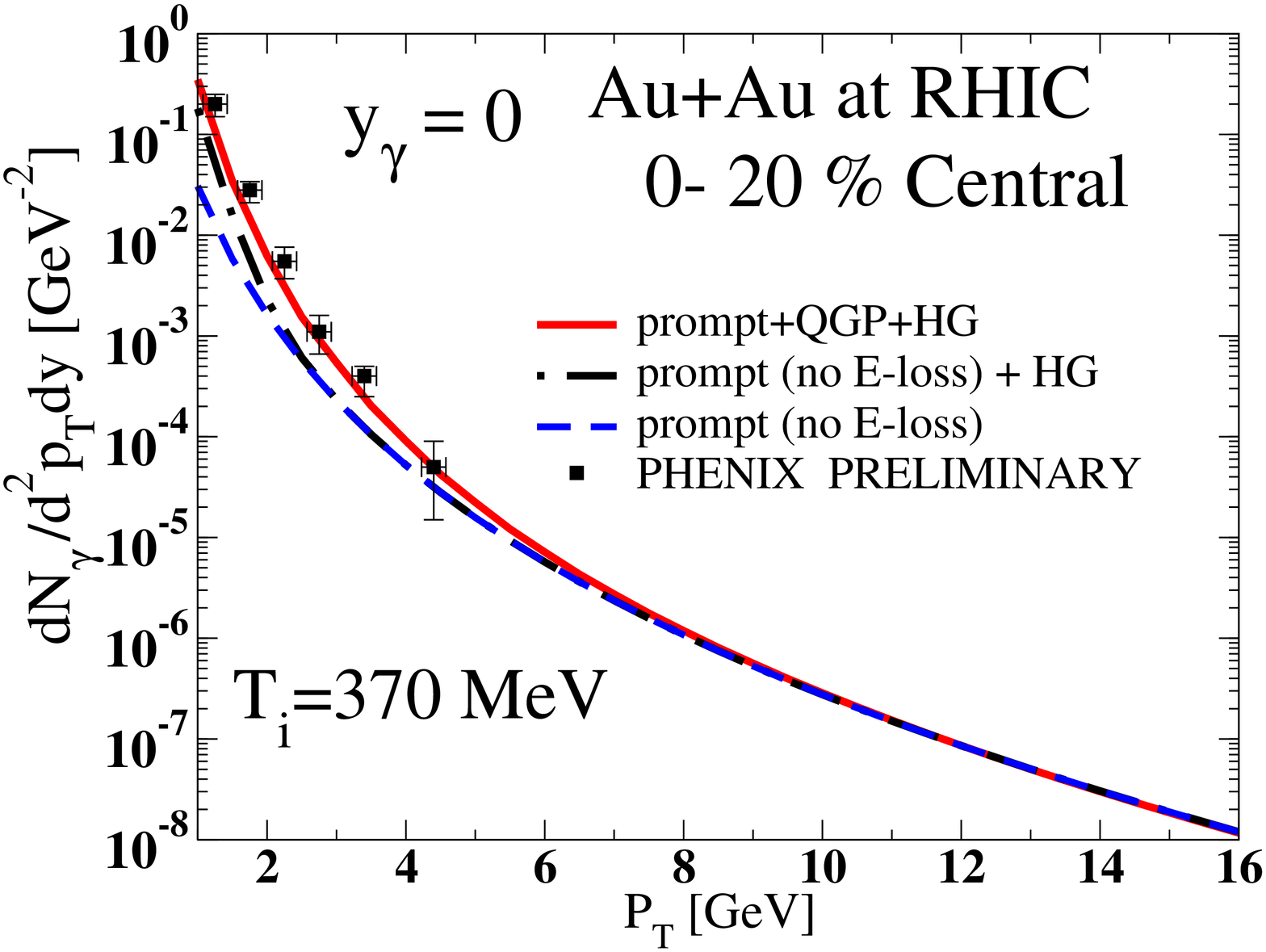}}
\vspace*{-1.0cm}
\caption{Left Panel: Contributing sources of high-$p_T$ photons at mid-rapidity in
central nuclear collisions at RHIC. Solid line: jet-photon conversion in 
the plasma; dotted line: bremsstrahlung from jets in the plasma; short dashed
line: thermal photons \cite{trg}; long dashed line: photons from fragmenting
jets; dot-dashed line: contribution from the primordial hard scattering.
Right panel: Total production of photons at RHIC, compared with new PHENIX data
\cite{phenix_photons}. The solid line represents all processes from the left
panel together with photons from the hadronic gas (HG) \cite{trg}; 
the dot-dashed line do not include any plasma-related contributions
but has those from the HG. 
Photons from pp collisions scaled to AA are shown by the dashed lines.  
\label{rhic_phot}}
\end{center}
\vspace*{-1.3cm}
\end{figure}

With the machinery at hand, the spectrum of photons produced in
nucleus-nucleus collisions can be calculated,
including those originating for jet-plasma interactions \cite{fms}, but 
consistently taking 
account energy-loss systematics. The sources include those active in pp
collisions: the direct photons produced by parton Compton and annihilation events, and
the fragmentation photons produced by bremsstrahlung from final state
partons. In AA collisions, the sources above still operate but the
fragmenting jets are now subject to energy-loss considerations. In addition,
hard partons traveling in the medium can also produce photons through
medium-induced bremsstrahlung. Finally, the conversion of leading partons to 
photons \cite{fms,tgjm} was found to be a significant contribution and should be treated
consistently with the other channels enumerated here. The different sources
for Au + Au collisions at RHIC are shown in Figure \ref{rhic_phot}. 
The agreement with the recent PHENIX photon data is very good. Furthermore, it is
satisfying to note that the modeling parameters were fixed prior to
its release \cite{tgjm}. Further note that when the jet-photon conversions
are omitted, the total photon production is reduced by up to 45\% around
$p_T$ = 3 GeV/c, showing the importance of this process. The total plasma
contribution appears important for $p_T <$ 6 GeV/c. Here also, the quality of
this data opens the door to additional investigations. Some have been done
\cite{dp} and more will follow, but importantly, electromagnetic
signals and hard hadronic probes are no longer disjoint observables.

\section{Conclusion}
It is not an overstatement to write that the measurement of 
electromagnetic probes in relativistic has produced some
very exciting results. All expectations
are that this will continue, with larger statistics samples and the
measurement of lepton pairs at RHIC. With the data shown at this meeting, we
are entering an era of precision measurements and modeling. 
For example, measurements of photon azimuthal asymmetry are now at reach. 

\section*{Acknowledgments}

I am grateful to all of my collaborators for their involvement with any aspect
of the work presented here. Of those, I also thank Simon Turbide for 
a critical reading of this manuscript. This work was supported in part by the
Natural Sciences and Engineering Research Council of Canada, and in part by
the Fonds Nature et Technologies of the Government of Quebec.

\end{document}